\documentclass[conference]{IEEEtran}

\usepackage{mathptmx} 

\usepackage{fancyhdr}
\usepackage[normalem]{ulem}
\usepackage[hyphens]{url}
\usepackage[sort,nocompress]{cite}
\usepackage[final]{microtype}
\usepackage[keeplastbox]{flushend}
\usepackage[bookmarks=true,breaklinks=true,colorlinks,linkcolor=black,citecolor=blue,urlcolor=black]{hyperref}
\usepackage{amsmath}
\usepackage{amssymb}
\usepackage{tabularx}

\usepackage{graphicx}
\usepackage{listings}
\usepackage{setspace}
\usepackage{subfigure}
\usepackage[most]{tcolorbox}
\usepackage{pifont}
\usepackage{array}
\usepackage{float}
\newcommand{\TODO}[1]{\textcolor{red}{TODO: #1}}

\newcommand{\TK}[1]{{{\color{red}TK: {#1}}}}

\newcommand{\insertFigure}[2]{
    \begin{figure}[t]
\setlength{\abovecaptionskip}{0pt}
\setlength{\belowcaptionskip}{0pt}
        \centering
        \includegraphics[width=\linewidth]{figs/#1.pdf}
	\vspace{-4mm}
        \caption{\small #2}
        \label{fig:#1}
    \end{figure}
}

\newcommand{\insertWideFigure}[2]{

    \begin{figure*}[h]
\setlength{\abovecaptionskip}{0pt}
\setlength{\belowcaptionskip}{0pt}
        \centering
        \includegraphics[width=\textwidth]{figs/#1.pdf}
	\vspace{-4mm}
        \caption{\small #2}
	\vspace{-2mm}
        \label{fig:#1}
    \end{figure*}

}

\newcommand{\squishlist}{
 \begin{list}{$\bullet$}
  { \setlength{\itemsep}{0pt}
     \setlength{\parsep}{3pt}
     \setlength{\topsep}{3pt}
     \setlength{\partopsep}{0pt}
     \setlength{\leftmargin}{1.5em}
     \setlength{\labelwidth}{1em}
     \setlength{\labelsep}{0.5em} } }

\newcommand{\squishlisttwo}{
 \begin{list}{$\bullet$}
  { \setlength{\itemsep}{0pt}
     \setlength{\parsep}{0pt}
    \setlength{\topsep}{0pt}
    \setlength{\partopsep}{0pt}
    \setlength{\leftmargin}{2em}
    \setlength{\labelwidth}{1.5em}
    \setlength{\labelsep}{0.5em} } }

\newcommand{\squishend}{
  \end{list}  }

\newcommand{\mys}{\textit{SAGAR}}    
\newcommand{\syscell}{\textit{systolic-cell}}
\newcommand{\ra}{\textit{RSA}}
\newcommand{\recnet}{\textsc{AdaptNet}} 
\newcommand{\core}{\textsc{AdaptNetX}} 
\newcommand{\revised}[1]{\textcolor{blue}{#1}}

\title{Learning Flexible GEMM Accelerator Configuration and Mapping-space using ML}
\author{
  Ananda Samajdar\\
  \textit{Georgia Tech}\\
  \textit{Atlanta, GA}\\
  \texttt{anandsamajdar@gatech.edu}
  \and
  Michael Pellauer\\
  \textit{NVIDIA}\\
  \textit{Boston, MA}\\
  \texttt{mpellauer@nvidia.com}
  \and
  Tushar Krishna\\
  \textit{Georgia Tech}\\
  \textit{Atlanta, GA}\\
  \texttt{tushar@ece.gatech.edu}
  
}

\begin{document}
\maketitle
\thispagestyle{plain}
\pagestyle{plain}


\begin{abstract}

The value of flexibility in Deep Learning accelerators to adapt to diverse layer shapes and sizes is well-understood.
Contemporary reconfigurable architectures depend on compilers or other components in the software stack for optimal configuration  and mapping search to fully exploit the benefits of flexibility. 
In this paper we show that the configuration and mapping space of flexible accelerators can be learnt using machine learning by casting it as a classification or recommendation problem.
The learnt model can be used to obtain the optimal configuration of the target accelerator in \textit{constant time without search}.
We propose \recnet{}, a recommender system for obtaining optimal configuration and mapping for GEMM workloads running on a \textsc{Reconfigurable Systolic Array} (\ra{}). 
%
\ra{} is designed to be configured such that it can operate across a spectrum from a single monolithic array to a distributed collection of smaller arrays of various sizes with flexible aspect ratios. 
This allows us to simultaneously achieve scalability and high mapping flexibility while preserving operand reuse.
\recnet{} demonstrates 95\% test accuracy 
 compared to an exhaustively searched optimal configuration, beating state-of-the-art classification techniques such as SVMs, XGBoost and MLPs.
We also present, \core{}, a specialized core to run \recnet{} in hardware.
Together, \ra{} and \core{} 
enable us to demonstrate a new class of flexible accelerators 
which  are capable of self-configuring in hardware for the given GEMM workload.
%
We present a 32.768 TOPS instance 
called \mys{} that is capable of providing the same mapping flexibility as a compute equivalent distributed system while achieving
3.5$\times$ more power efficiency and
3.2$\times$ higher compute density demonstrated via architectural and post-layout simulation.

\end{abstract}
\section{Introduction}
\label{sec:introduction}

\insertFigure{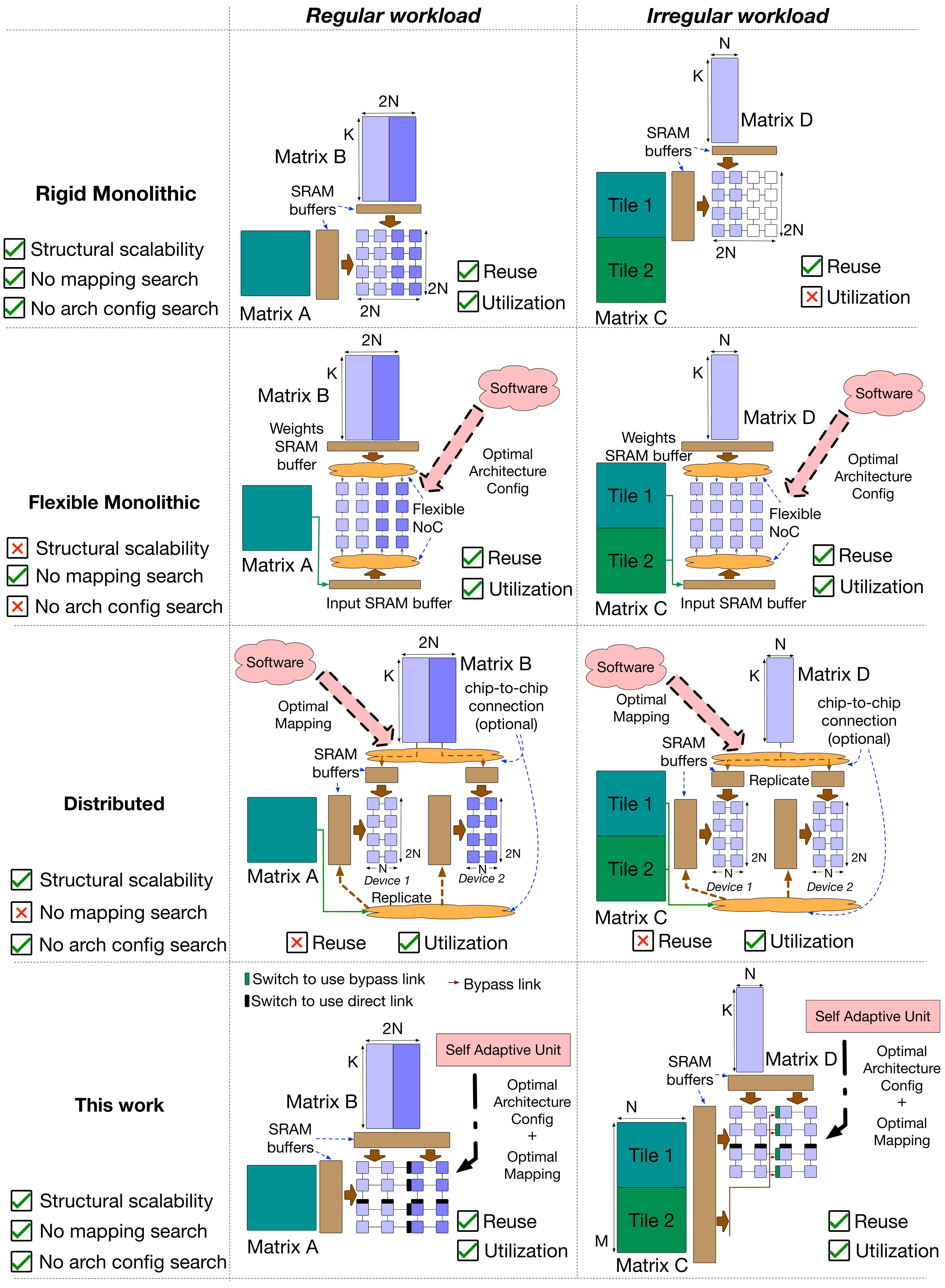}{Comparison of scalability, utilization, and operand reuse in traditional monolithic and distributed accelerators, and the position of the proposed architecture}

\insertFigure{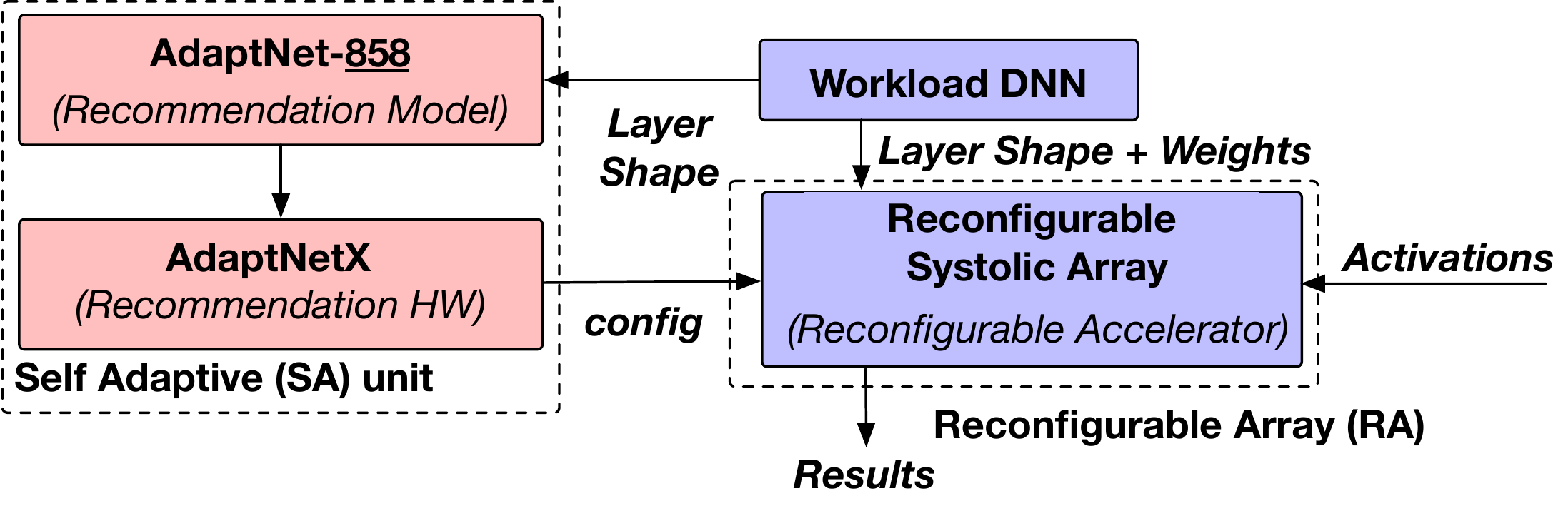}{
    The constitution and interactions of the self adaptive (SA) and reconfigurable array (RA) components to make up the SARA accelerator called \mys{} in this work. 
}

Custom architecture design enables us to achieve high performance and energy efficiency for a given class of workloads in post Moore's Law era.
Highly specialized architectures however are inflexible to any variation in the nature of workload and thus can easily be rendered obsolete.
To mitigate this limitation, there has been an increasing interest in developing flexible architectures which have additional components (interconnects, buffers, and configuration registers) to support changing workload requirements.
%
For popular applications like DNN acceleration, several such flexible architectures have been proposed \cite{brainwave, zhang-fpga-2015, tetris, tangram, maeri}.

In all of the prior works on flexible DNN accelerators, however, 
the onus of finding and setting the best configuration lies on the software stack, typically using a compiler/mapper~\cite{gamma, timeloop, mindmappings, zhao2019mrna}. 
This dependence causes a few deployment challenges: (i) a cost model has be to developed and integrated as an optimizer into the compilation stack to help find optimal mappings, without which the flexible design loses utility,
(ii) an expensive configuration and mapping search has to be performed at compile-time before scheduling any workload. 
Usually mapping search in software is performed via exhaustive, heuristic or optimization algorithm-based approaches which take about a few seconds to hours~\cite{timeloop, gamma, zhao2019mrna}, even with sophisticated ML assisted frameworks like autoTVM \cite{autotvm}.
(iii) the search-time overhead also eliminates opportunities for deploying such flexible accelerators for domains/applications with soft or hard-real time inference targets.

In this work, we demonstrate that the mapping and configuration space of a reconfigurable accelerator can be \textit{learnt} by a machine learning (ML) model,
which can then be used to query for optimal parameters for any workload at constant time.
Dependence on software stack can be eliminated by incorporating this learnt model into the hardware itself and querying it in runtime.
We illustrate this via two contributions:

\textbf{\textit{First,}} we design a scalable reconfigurable hardware optimized for GEMM workloads called \textsc{Reconfigurable Systolic Array} (\ra{}).
\ra{} is developed upon the intuition that flexible accelerators often need to trade-off utilization, data reuse, and hardware 
complexity (i.e., scalability). 
This is illustrated in \autoref{fig:scale-out-example-intro-new}.
\textit{Rigid Monolithic} arrays (e.g., TPU's systolic array~\cite{tpu}), 
are simple to construct but offer no flexibility leading to high under -utilization for many workloads~\cite{sigma, scalesim-ispass}.
\textit{Flexible Monolithic} arrays (e.g., MAERI~\cite{maeri}, Eyeriss\_v2~\cite{eyerissv2}, SIGMA~\cite{sigma}) provide flexbility via clever use of interconnects and configuration logic, enabling high utilization for a majority of workloads.
However, the increased hardware complexity hinders scaling,
and the design requires external software support to exploit the benefits of reconfigurability~\cite{timeloop, mindmappings, gamma}.
Distributed architectures (e.g., Tangram~\cite{tangram}, Simba~\cite{simba}) 
help address the utilization challenge, since irregular workloads can be tiled on to these smaller arrays.
However, this architecture leads to loss in  spatial reuse (i.e., direct data-forwarding) that monolithic designs provide, 
and also requires data replication across the SRAMs of the individual arrays.
Data replication leads to a decrease in overall on-chip storage capacity, leading to a loss of temporal reuse due to smaller tiles.
Moreover, distributed arrays can exacerbate the mapping search problem~\cite{simba, timeloop}.
\ra{} aims to address the shortcomings of all three design strategies. It is a flexible accelerator capable of supporting mappings that can be realized by monolithic as well as distributed arrays by configuring to variable array dimensions and number of sub-arrays (as depicted later in \autoref{fig:high-bw-smart-systolic-array}(d)), thereby enhancing both utilization and reuse. 
In practice, \ra~closely approximates a flexible monolithic design, with a fraction of area cost.

\textbf{\textit{Second,}} we present a systematic mechanism to cast the architecture configuration as a ML classification problem and discuss considerations for optimal model design, training, and performance of the model at inference. Specifically, we develop a custom ML recommendation system model called \recnet{} that achieves a recommendation accuracy of 95\% on a dataset of 200K GEMM workloads, and on average(GeoMean) 99.93\% of the best attainable performance (Oracle). 
We also  design a custom hardware unit to run \recnet{} called \core{}.
\core{} enables to get a recommendation response for any query in about 600 cycles which is at least about 6 orders of magnitude faster than software.
Furthermore, \core{} consumes the same hardware real-estate and roughly the same on-chip memory capacity
\footnote{The only change in \recnet{} between various \ra{} is the weight of the output layer, which is small in comparison to the embedding table which takes most of the on-chip space}
for different arrays, thus proving to be a scalable solution in contrast to approaches like using configuration caches.
With \core{} the configuration lookup using \recnet{} can be performed at runtime, without involving the software stack.

Together, these two components enable us to develop 
a new class of accelerators that we call \textit{Self Adaptive Reconfigurable Array (SARA)} (\autoref{fig:sara-concept}). 
SARA accelerators can self adapt at runtime to optimized configurations for the target workload, without 
requiring compile-time analysis.
We demonstrate an instance of SARA that we name `Shape Adaptive GEMM AcceleratoR (\mys{}
\footnote{means Ocean in Sanskrit, reflecting the shape flexibility of our accelerator.})
as shown in \autoref{fig:sara-concept}
and evaluate its performance across various 
configurations.
We show that \mys{} has 
3.2$\times$ higher compute density and
3.5$\times$ improved power efficiency, over equivalent scaled-out systolic array.
The extra flexibility costs $<$10\% in area and 
50\% in power, compared to equivalent scaled-up systolic array.
Compared to an area normalized state-of-the-art flexible scalable accelerator \cite{sigma}, \mys~ incorporates 
45\% more compute.
When comparing compute-equivalent configurations,
\mys~ consumes 43\% less power and 30\% less area.

\vspace{-2mm}
\section{Reconfigurable array design}
\label{sec:arch}


\insertFigure{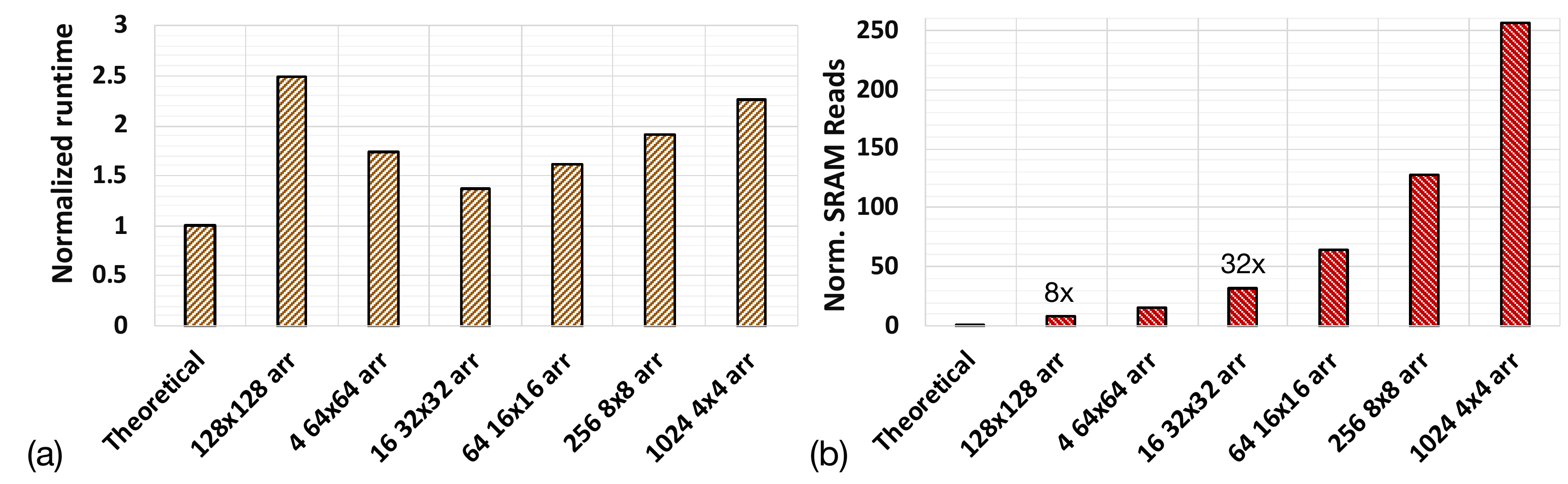}{
    The trade-off between improved runtime and lost operand reuse in compute equivalent monolithic and distributed systolic array configurations.
    (a) the theoretical minimum runtime, and the runtime obtained for stall free operation of monolithic and compute normalized distributed systolic array settings; and 
    (b) the corresponding SRAM reads, normalized to theoretical minimum reads required when multiplying a $256\times64$ matrix with another $64\times256$ matrix. 
}
\vspace{-2mm}
\subsection{Motivation}
\label{subsec:motivation}

To help understand the trade offs involved in choosing a performant configuration, and the associated loss of reuse we perform a simple experiment.
We run one GEMM operation, involving operand matrices of sizes sizes $256 \times 64$ and $64 \times 256$ on various systolic array configurations. 
These are, a $128 \times 128$ monolithic array, and five distributed scale-out configurations viz. 4 $64 \times 64$ arrays, 16 $32 \times 32$ arrays, 64 $16 \times 16$ arrays, 256 $8 \times 8$ arrays, and 1024 $4 \times 4$ arrays.
We obtain the runtime and memory accesses for running this workload on all the array configurations using SCALE-Sim \cite{scalesim-arxiv} (see \autoref{subsec:sim-eval}).
In \autoref{fig:motivation-data-scaleup-scaleout}(a) we show the runtime normalized to the theoretical minimum cycles required.
Please note that with the chosen matrix dimensions, the systolic arrays in all the configurations are mapped 100\% with useful computation.
The differences in runtime in various arrays under 100\% mapping efficiency is attributed to the array filling and draining at each serialization step (see sec III in \cite{scalesim-ispass}).
We observe that the configuration with $32\times32$ array is the most performant, beating the monolithic configuration by about 2$\times$. 
In \autoref{fig:motivation-data-scaleup-scaleout}(b) we depict the SRAM read accesses performed by all the array configurations, normalized 
to the theoretical minimum number of reads possible.
From this figure we observe that the $32\times32$ configuration performs about 4$\times$ more memory accesses then the monolithic.
The excess memory accesses, which lead to reduced energy efficiency, result from the loss of wire reuse.

From the discussion above we make two observations.\\
(i) Distributed arrays are more performant than an equivalent monolithic array, even when mapping efficiency is 100\% on both.
However, the optimal size of each device in a distributed setting is workload dependent.
(ii) Monolithic configurations are strictly more energy efficient than distributed arrays, due to loss the of spatio-temporal reuse in the latter.

\newcommand{\multarrdim}[0]{\begin{tabular}{c} Multi-array \\ Mapping \end{tabular}}
\newcommand{\vararrdim}[0]{\begin{tabular}{c} Variable \\ Dimensions \end{tabular}}
\newcommand{\selfconf}[0]{\begin{tabular}{c} Self \\ Configurable \end{tabular}}

\begin{table}[t] 
\centering 
\setlength{\abovecaptionskip}{3pt}
\setlength{\belowcaptionskip}{0pt} 
\caption{\small Previous accelerator proposals categorized in terms of computation support, and flexibility of hardware and mapping. 
Accelerators are categorized into various types introduced in \autoref{fig:scale-out-example-intro-new} viz. Rigid Monolithic (RM), Flexible Monolithic (FM), and Distributed (Dist)} 
\resizebox{\linewidth}{!}{
\Large
\begin{tabular}{l|c|cc|ccc|c}

\hline
 && \multicolumn{2}{c}{Mapping Capability} & \multicolumn{3}{c}{Flexibility} & \\
 & Type & Homogenous & Heterogenous & Dataflow & \vararrdim & \multarrdim & \selfconf \\
 
 \hline
 \vspace{1mm}
 Zhang et al. \cite{zhang-fpga-2015} & FM &
 \checkmark && 
 \checkmark & & &\\
 
 \vspace{1mm}
 Eyeriss \cite{eyeriss} & RM &
 \checkmark& &
 & & & \\
 
 \vspace{1mm}
 Alwani et al. \cite{alwani-2016-MICRO-fused} & RM &
 & \checkmark &
 & & & \\

 \vspace{1mm}
 NeuroCube \cite{neurocube} & Dist &
 & \checkmark &
 & & \checkmark & \\
 
\vspace{1mm}
MAERI \cite{maeri} & FM &
\checkmark & \checkmark&
& \checkmark & & \\

\vspace{1mm}
TPU \cite{tpu} & RM &
\checkmark & &
& & & \\

\vspace{1mm}
Flexflow \cite{flexflow} & FM &
\checkmark &&
\checkmark & & & \\

\vspace{1mm}
Tetris \cite{tetris} & Dist &
& \checkmark &
& & \checkmark & \\

\vspace{1mm}
Brainwave \cite{brainwave} & Dist &
&\checkmark & 
& &\checkmark & \\

\vspace{1mm}
Simba \cite{simba} & Dist &
&\checkmark &
& & \checkmark & \\

\vspace{1mm}
Tangram \cite{tangram} & Dist &
& \checkmark &
& & \checkmark &\\

\vspace{1mm}
Cascades \cite{cascades} & FM &
& \checkmark &
\checkmark  & & \\

\vspace{1mm}
Sigma \cite{sigma} & FM &
& \checkmark &
& & & \\

\vspace{1mm}
Planaria \cite{planaria} & FM &
& \checkmark &
&\checkmark & & \\

\vspace{1mm}
\textbf{\mys{} (This work)} & &
\checkmark & \checkmark &
\checkmark & \checkmark & \checkmark & \checkmark \\

\hline 

\end{tabular}
}
\label{table:related-works} 
\end{table}

In \autoref{table:related-works} we inspect a few well known accelerator proposals in terms of scalability and potential to maximize utilization. 
We notice that simple architectures that are easy to scale in size, under perform on extracting operand reuse. On the other hand, architectures with sufficient flexibility are not scalable. 
None of the architectures, including the ones with multiple arrays and NoC support, can create variable sized arrays or flexible array dimensions which can help simultaneously achieve high mapping efficiency and data reuse.

In the next subsections we develop a flexible design obtained by augmenting a base monolithic systolic array with additional bypass paths along the row and columns.
This design enables use to chose between configuration akin to a large single array or a collection of smaller arrays, whenever required.

\vspace{-2mm}
\subsection{Compute array}
\label{subsec:bypass-only-array}

\textbf{Traditional MAC units.} 
In \autoref{fig:traditional-mac-array}(a) we show a traditional systolic array constructed by laying down Multiply-and-Accumulate (MAC) (\autoref{fig:traditional-mac-array}(b)) units in a 2D grid.
Each MAC unit is designed to get an operand from either both (\textit{Left in, Top in}) ports or from either of the ports, and perform multiplication and addition operation.
In the next cycle the operand data received is sent to its neighbour over the peer-to-peer links.
The internal registers, and multiplexers enable the array to work in output stationary (OS), weight stationary (WS), and input stationary (IS) modes of operation \cite{eyeriss}.
This simple mechanism of data movement results in high wire reuse, but at the same time restricts the mapping of compute only to those operations which require same set of operands to be mapped along a row or a column.

\insertFigure{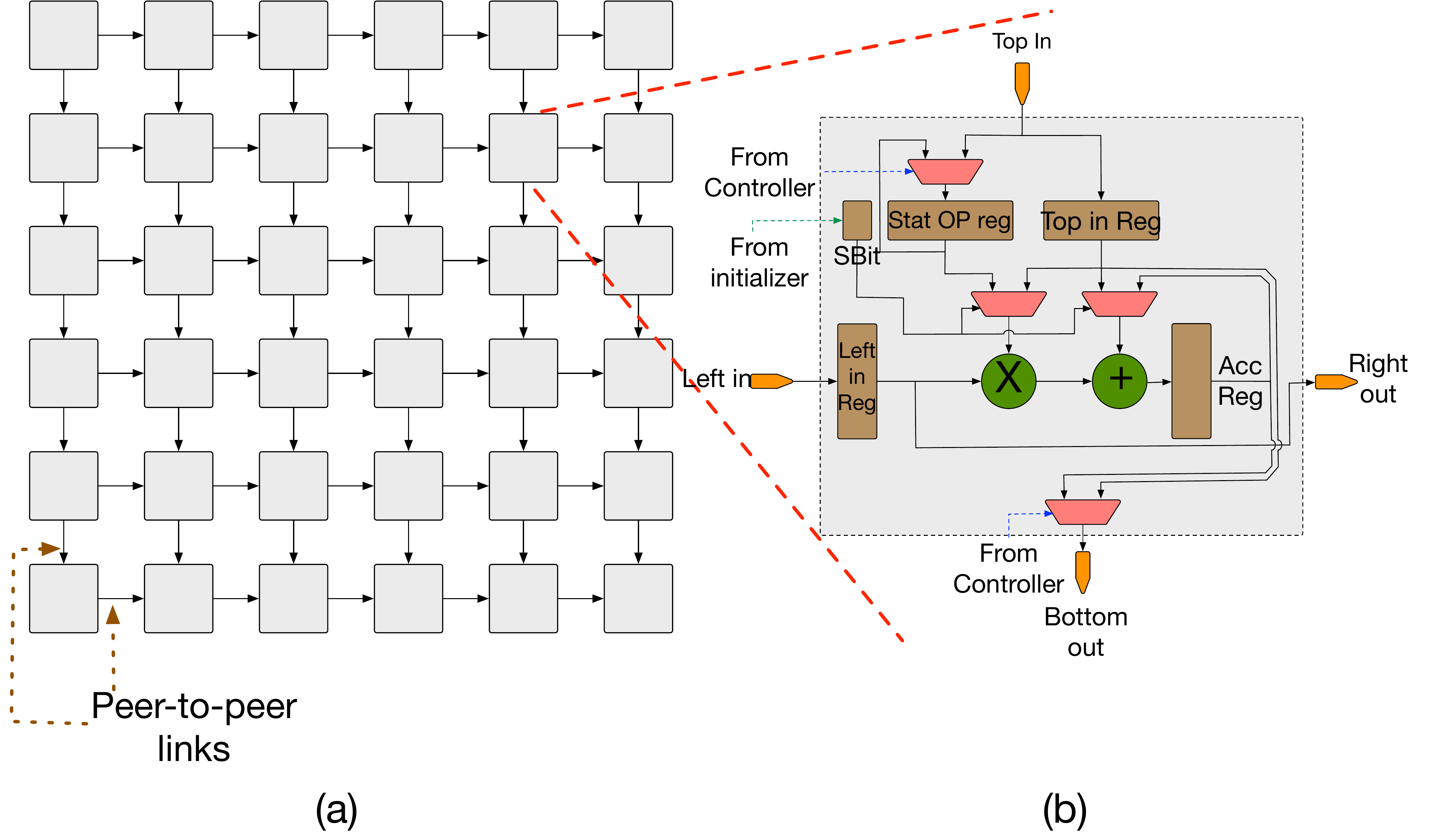}{
    (a) A systolic array of traditional MAC units, 
    (b) the architecture of a traditional MAC unit
}

\insertWideFigure{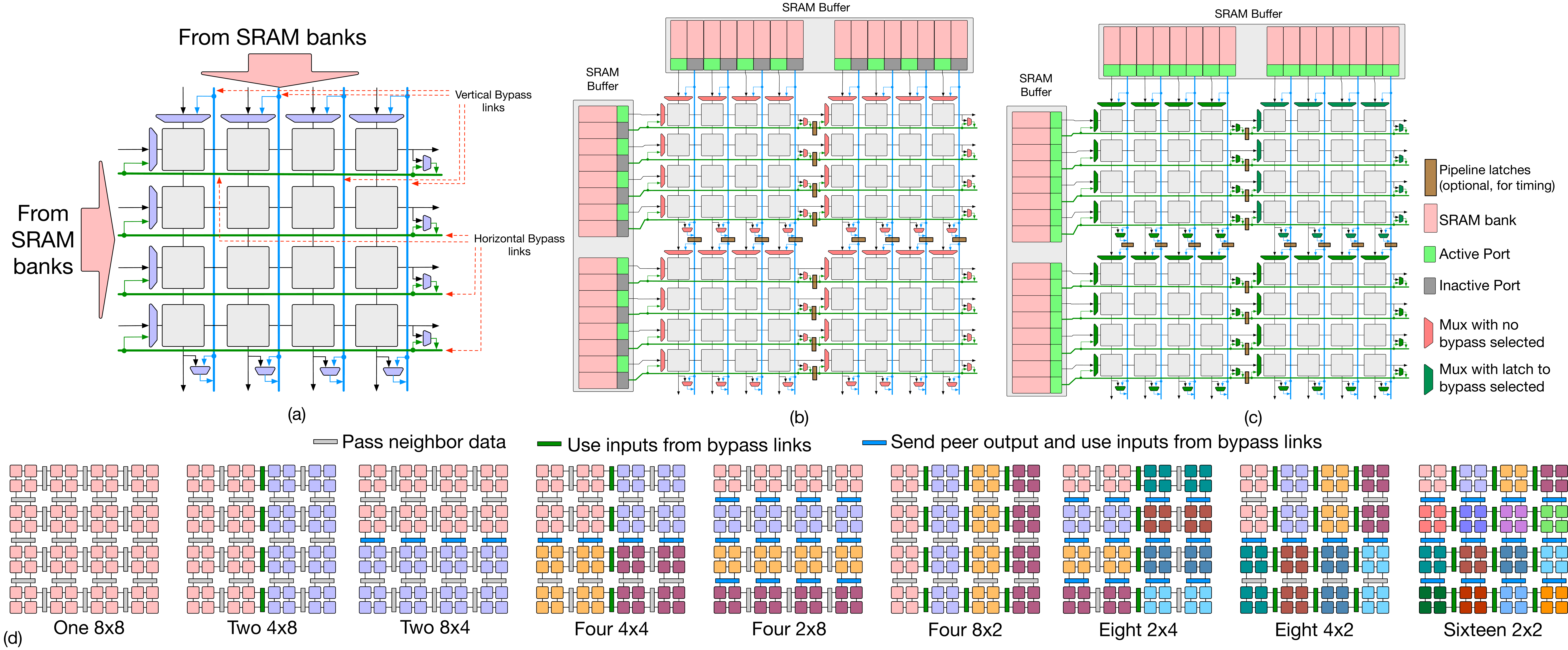}{
    (a) Construction of a $4\times4$ \textit{systolic-cell} with bypass muxes and bypass links.
    (b) A $8\times8$ SMART systolic array operating in scale-up configuration. Each $4\times4$ \textit{systolic-cell} is connected to its neighbor with the peer-to-peer links as the bypass muxes are turned off. The SRAM ports connected to bypass links are unused. 
    (c) Configuration of bypass muxes to enable the  $8\times8$ SMART systolic array to work as a scaled-out distributed collection of systolic arrays. The bypass muxes are turned on to allow systolic-cells to directly connect to the SRAM ports which are all active.
    (d) Possible monolithic and distributed configurations possible in the reconfigurable Smart-Systolic Array(SSA) using $2\times2$ \syscell s
}


\textbf{Systolic Cells.}
The mapping flexibility in systolic arrays can be improved by allowing adjacent MAC units to work on different operands. 
To enable this, the architecture needs to provision for additional links from the SRAM to the MACs. 
Providing such links to each MAC however is costly in terms of area as well as energy since the spatial reuse over wires is compromised. 
To simultaneously achieve mapping flexibility and the advantages of spatial reuse in systolic arrays, we propose a design called \syscell. 
A \syscell~ is a small grid of traditional MAC units augmented with multiplexers at the edges. 
This enables them to chose the operands from the neighbouring MAC units or a separate set of operands available via bypass links.
The MACs within a \syscell~ are connected using peer-to-peer links similar to that of a traditional systolic array.
\autoref{fig:high-bw-smart-systolic-array}(a) shows a $4\times4$ \syscell~ example. 
Please note that choice of the size of a \syscell~ is implementation dependent.
In general, the smaller the cell size, higher the mapping flexibility, which comes at a cost of slightly increased area and power.

\textbf{Scale-up and Scale-out using Systolic Cells.} 
Larger arrays can be created by arranging and connecting the \syscell s as depicted in \autoref{fig:high-bw-smart-systolic-array}(b) using the peer-to-peer links.
At the edge of each \syscell ~the muxes can be configured to connect to the bypass links. 
Please note that dedicated bypass links are allocated to each \syscell{} to allow concurrency.
Attaining flexible mapping in such a design is a matter of configuring the multiplexers of the \syscell s.
Depending on the mapping , an user can chose not to use the bypass paths at all and use the entire array as a single monolithic unit by setting the multiplexers to accept data only to/from the peer-to-peer links, (this is the case depicted in \autoref{fig:high-bw-smart-systolic-array}(b)), which is equivalent to a \emph{scaled-up} configuration.
One the other hand, the user can set all the multiplexers to accept and deliver data solely to the bypass links, therefore operating as a cluster of arrays, each the size of a \syscell.
This configuration, depicted in \autoref{fig:high-bw-smart-systolic-array}(c)  is equivalent to a \emph{scaled-out} configuration.
\autoref{fig:high-bw-smart-systolic-array}(d) illustrates some of the possible configurations constructed using a 64 MAC units with $2\times2$ \syscell s.
As can be observed in this figure, not only can the array be configured to work in fully monolithic or fully distributed configurations, but also in any of the configurations in between.
By setting the appropriate muxes in either pass-through or bypass modes, sub-arrays larger than \syscell ~size can be constructed (eg. $4\times4$,  $8\times4$ etc in this example). 
Each of the sub-arrays have access to the scratchpad memory using the bypass links.
Please note that when fully utilized, a larger systolic array improves energy efficiency over a distributed configuration of same number of MAC units by exploiting wire reuse and reducing SRAM reads.
The availability of such variety of choices for reconfiguration leads to flexible and efficient mapping, hence improving the utilization and energy efficiency of the design.

\vspace{-2mm}
\subsection{Bypass links}
\label{subsec:bypasslinks}


Adding a dedicated bypass link from the SRAM bank to each \syscell~ along that row/column is necessary to attain full throughput from the array.
Given the nature of the data movements in systolic arrays, we recognize that the vertical links can be used for both second input and the output operands. 
In \autoref{table:wire-scaling}, we examine the bandwidth requirements from the bypass links for the three systolic dataflows in a distributed setting, by contrasting it to the requirements of the operands. 
These requirements clearly dictate that high bandwidth bypass are necessary.
Another addition in our proposed architecture are the switches at the edges of the \syscell{}s. 
However, these switches are simple multiplexers, which are configured statically for a given workload, without the need for any additional logic.

\textbf{Scalability via Pipelining.}
On-chip wire scalability studies such as SMART\cite{smart} have shown that it is possible to traverse a few millimeters (9mm to 11mm) of wire length in 1ns before latching the signal. 
The authors in SMART achieved this using conventional asynchronous repeaters (a pair of inverters) placed 1mm apart. 
In \ra{}, repeated wires offer an opportunity to not only cross a single-systolic cell in a cycle, but in fact bypass multiple systolic cells within a single-cycle. 
In our reference architecture \mys{}, we perform place-and-route to determine the number of systolic cells per pipeline stage of the bypass links. 
At 28nm, we find that 8 systolic cells can be bypassed at 1GHz, as we demonstrate later in \autoref{subsec:eval-impl}, \autoref{fig:rebuttal_pnr_chart}(h).
Note that pipelining the bypass links only adds a few cycles of fill time to the \ra{}, and does not impact the internal timing of the systolic array within each systolic cell (which is itself pipelined at each MAC unit).

\newcommand{\inputs}[0]{\begin{tabular}{c} Input Mat1 \\ (Activations) \end{tabular}}
\newcommand{\weights}[0]{\begin{tabular}{c} Input Mat2 \\ (Filters) \end{tabular}}
\newcommand{\os}[0]{\begin{tabular}{c} Output \\ Stationary \end{tabular}}
\newcommand{\ws}[0]{\begin{tabular}{c} Weight \\ Stationary \end{tabular}}
\newcommand{\is}[0]{\begin{tabular}{c} Input \\ Stationary \end{tabular}}
\newcommand{\hiI}[0]{\begin{tabular}{c} High \\ (Inputs) \end{tabular}}
\newcommand{\hiW}[0]{\begin{tabular}{c} High \\ (Filters) \end{tabular}}
\newcommand{\hiO}[0]{\begin{tabular}{c} High \\ (Outputs) \end{tabular}}

\begin{table}[t]
\Large
\centering 
\setlength{\abovecaptionskip}{3pt}
\setlength{\belowcaptionskip}{0pt} 
\caption{\small Bandwidth requirements for the bypass links for various dataflows, contrasted to the requirements of operands (names in parenthesis reflects the corresponding operands in 2D convolutions) }
\resizebox{\linewidth}{!}{
\begin{tabular}{lccccc}

\hline
    &\multicolumn{3}{c}{Operands} & \multicolumn{2}{c}{Links} \\
    & \inputs & \weights & Outputs  & Hor. Bypass & Ver. Bypass \\
\hline
\os & High & High & Low  & \hiI & \hiW \\
\hline
\ws & High & Low  & High & \hiI & \hiO \\
\hline
\is & Low  & High & High & \hiW & \hiO \\
\hline

\end{tabular}
}
\label{table:wire-scaling} 
\end{table}
\vspace{-2mm}
\subsection{Scratch pad memory}
\vspace{-1mm}
The array constructed from \textit{systolic-cells} is backed by SRAM scratchpad memories, which are constructed as two individual buffers.
Each of these buffers is dedicated to one of the operand matrices. 
Such scratchpad SRAM buffers are common in accelerators, and are designed to reduce the number of off chip accesses and facilitate temporal reuse. 
Each operand buffer is operated in a double buffered fashion, so that the prefetch latency can be minimized. 
The system also contains a third buffer which is used to store generated outputs elements.

To support the bandwidth of bypass links, we provision for this extra bandwidth by increasing the number of memory banks in the scratchpad SRAM buffers.
%
%
Despite having the same number of SRAM ports as in a distributed configuration, this approach has a couple of advantages over the latter.
\textit{First}, there is no replication of data required, which otherwise reduces the effective capacity of the system therefore adversely affecting reuse.
In our design by eliminating replication we inherently improve the temporal reuse of operands. 
\textit{Second}, each of the \textit{systolic-cells} can access data in the entire operand buffer. Due to unified memory control of each buffer, operation like multi-cast are implicit in form of read collation, which improves energy efficiency without impacting performance. 
We describe the impact on reads and energy efficiency in detail in \autoref{subsec:sim-eval}.

\vspace{-1mm}
\subsection{Control}
\vspace{-1mm}

\insertFigure{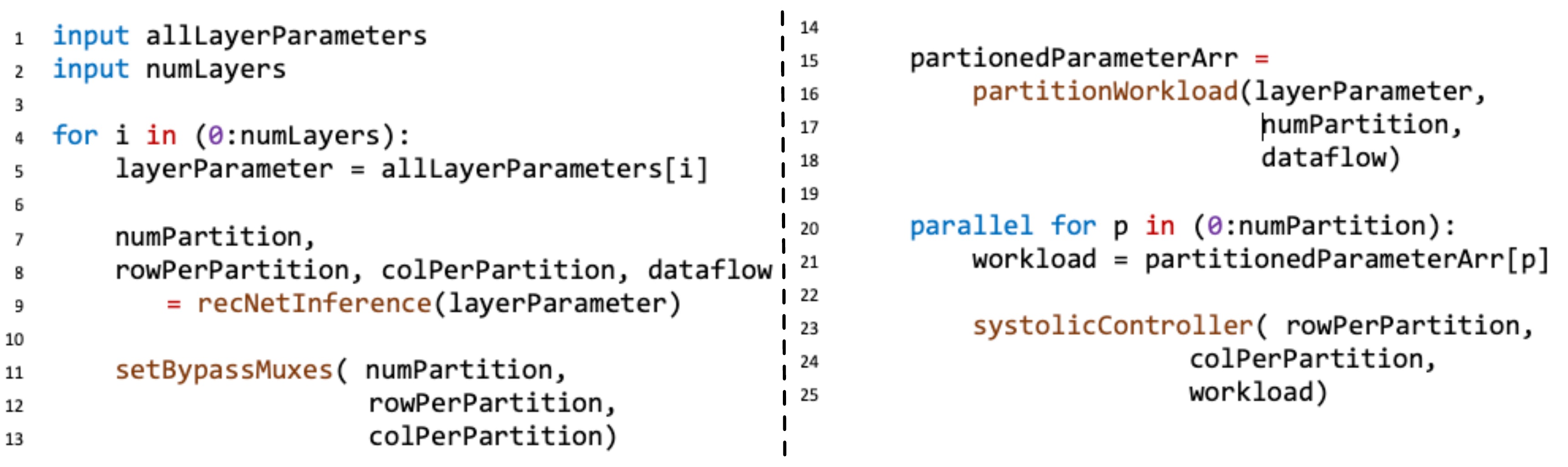}{Psuedocode depicting the control logic}


\autoref{fig:pseudocode} shows the control logic executed for each GEMM workload or DNN layer.
The following steps describe the logic. \\
\textbf{1.} \textcolor{brown}{recNetInference}(): In this work we use a recommendation system based described in \autoref{sec:airchitect}. The model takes in the layer parameters and recommends a configuration, which is the most efficient for the workload.
\textbf{2.} \textcolor{brown}{setBypassMuxes}(): Next, the bypass muxes are set in the compute hardware to realize the partitioned configuration. 
This is accomplished by writing select values to a register, whose individual bits drives the select lines. These configurations stay static throughout the GEMM computation.
\textbf{3.} \textcolor{brown}{partitionWorkload}(): The control logic, then partitions the original workload by marking portions of the original operand arrays to be used by each individual partition.
\textbf{4.} \textcolor{brown}{systolicController}(): Finally, for each partition, an instance of systolic array controller is initiated to drive the GEMM operations to completion and orchestrate the required data movement. Please note that in contrast to a traditional systolic array like TPU, multiple control units are required to work in parallel.

\vspace{-2mm}
\section{Recommendation Model}
\label{sec:airchitect}

   
\insertWideFigure{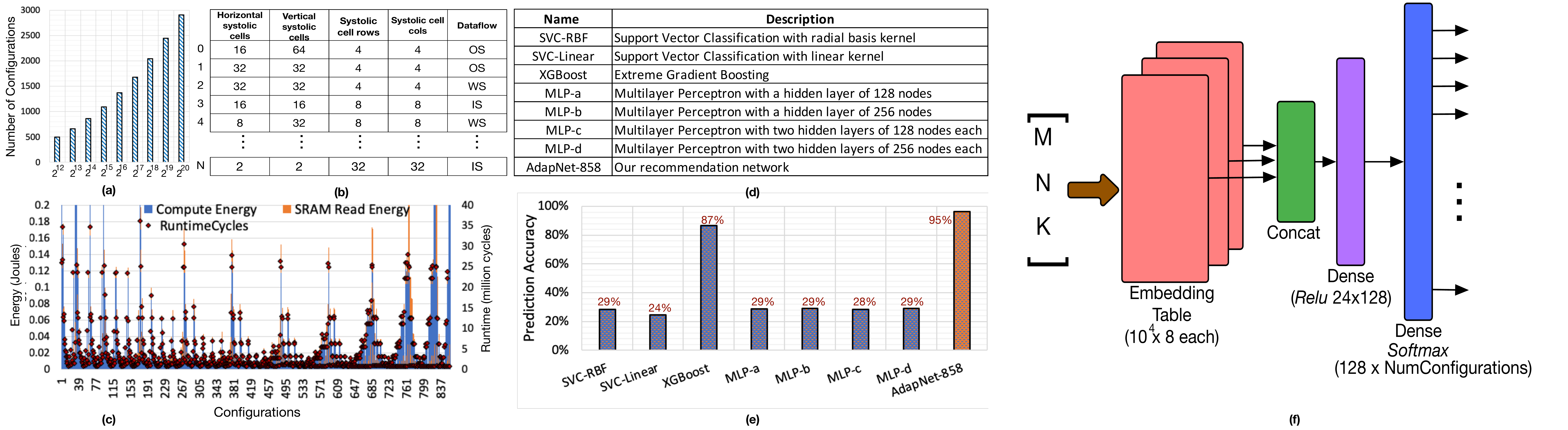}{
    (a) Size of the configuration space wrt number of MAC units for a \syscell~ based flexible array
    (b) Example of configurations predicted by \recnet~ indexed by category ID
    (c) Performance and energy consumption by various configurations when running layer 19 of FasterRCNN
    (d) Chart showing the name and description and name of the various classifiers used in this work
    (e) The accuracies obtained by classifiers on predicting the architecture parameters for out dataset of RSA configurations with $2^{14}$ MAC units
    (f) Architecture of our proposed recommendation network
}

This section describes a neural network based recommendation unit which can simultaneously predict the optimal architecture configuration and  mapping strategy, when a workload arrives. 
This system solves two problems.
First it minimizes the changes required in a compiler for configuration and mapping search, thus easing deployment. 
Second, it enables real time reconfigurability.
Given that a large reconfigurable array is most likely be deployed in data-center like use cases, the capability to adapt in real time will help achieve improved resource allocation and consequently meeting tight service-level-agreements (SLA).

\vspace{-2mm}
\subsection{Architecture design as ML problem}
To facilitate learning the design space we have to frame the search problem into a ML task framework like classification or regression. 
We found that framing this as a classification or recommendation task works the best.  
This abstraction lets us leverage the existing works and models which have been invented by the ML community. 
An important step in solving this problem is to define the output space of the model.
It is natural to assign bins for the each of the design parameters and independently predict the optimal values for each parameter of interest. 
However, this would require a separate model to be trained and queried for each design parameter.
We show that multiple parameters can be combined into a single output class and consequently can capture the design space using a single model.

In our case, the output space comprises of (i) The number and logical layout of the partitions, (ii) The dimensions of the arrays in each partition, and (iii) the mapping/dataflow to be used eg. output stationary (OS), weight stationary (WS),and input stationary (IS).
\autoref{fig:adaptnet-config-arch-wide}(b) shows this output space captured as categories of architecture configurations, indexed by the class ID.
The learned classifier is expected to select an architecture configuration and corresponding mapping strategy which provides the optimal performance for the workload. 
To better visualize the complexity of the design space, in \autoref{fig:adaptnet-config-arch-wide}(c) we depict the runtime and energy consumption for computation and SRAM reads when running layer 19 of FasterRCNN (see \autoref{subsec:sim-eval}) for the different architecture configurations and dataflows.
We observe that determining the optima is a non-trivial task; and the likelihood to chose a sub-optimal configuration is high, when naive methods are used, resulting in significant performance and energy costs.
Moreover, the best configuration for this layer is using 256 partitions laid out as a $8\times32$ grid of $16\times4$ arrays using WS dataflow, which does not conform to conventional practice of using square or near-square layouts.


\insertFigure{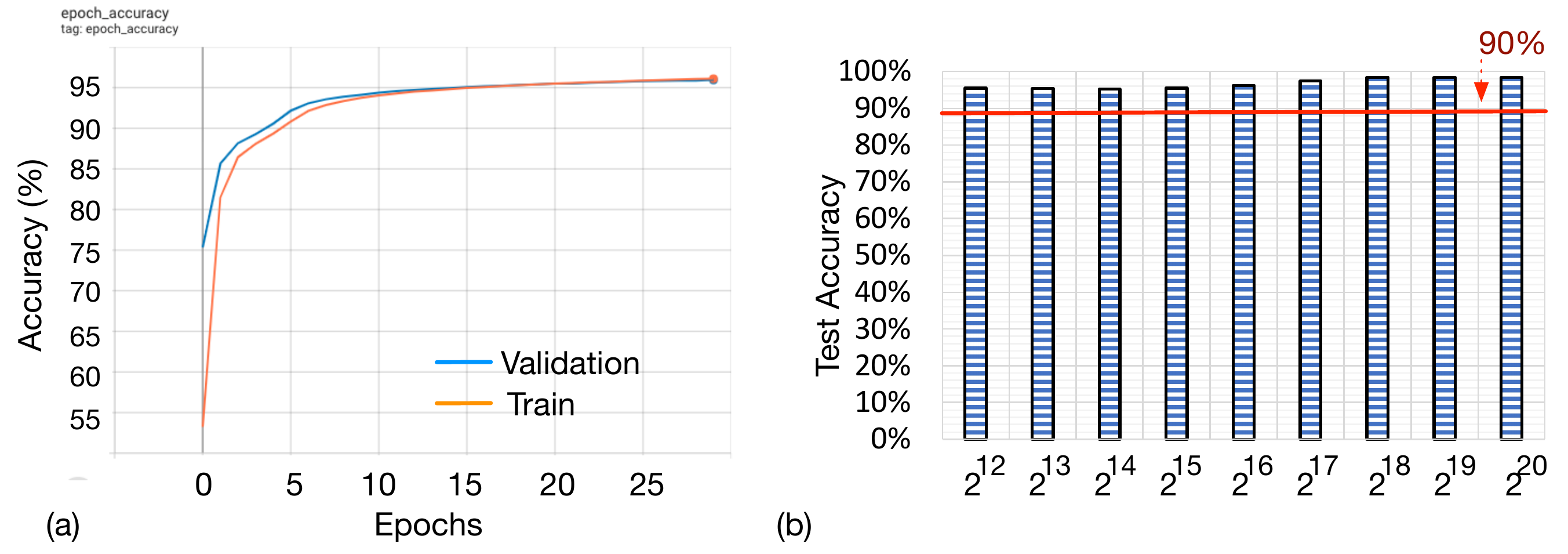}{
    (a) The training and validation accuracies obtained during the training step in \recnet{}
    (b) Test accuracies obtained on test sets for \recnet s trained on \syscell{} based flexible array with various MAC units
}

\subsection{Recommendation Neural Network}
\textbf{Dataset generation.} We generated a dataset of about 2 million workloads, by sampling M, N, and K dimensions from a uniform distribution of positive integers $<= 10^4$. For each such workload dimension we searched through the configuration space of the reconfigurable array design using $2^{12}$ MAC units to find the optimal (minimum runtime) configuration using SCALE-Sim simulator, modified for fast runtime estimation. 
When using a server cluster with about 200 Xeon cores, it takes about a week to obtain all the samples.

\textbf{Choosing the classifier.} Abstracting the problem in the form of a classification naturally opens up the choice of using existing classification algorithms. 
We explored a handful of pre-existing classifiers, some of which are listed in \autoref{fig:adaptnet-config-arch-wide}(d).
The Support Vector Classifiers and the XGBoost models we use are standard implementations provided in scikit-learn~\cite{sklearn} and xgboost\cite{xgboost} python packages respectively. 
We implement the MLPs in keras subpackage in tensorflow and train them for 20 epochs.
In \autoref{fig:adaptnet-config-arch-wide}(e), we show the prediction accuracy of these models on a test set of 200K points, after the model has been trained on 90\% of the dataset. 
It is interesting to observe that among all the models only XGBoost was able to reasonably learn the design space and achieve about 87\% prediction accuracy. 

\textbf{Recommendation Model:} The performance of XGBoost model is encouraging and demonstrates that the design space can be learnt. 
To further improve the prediction performance of the model, we hand designed a recommendation neural network. 
We take inspiration from typical neural network based recommendation systems like DLRM\cite{dlrm}, which is constructed by augmenting embedding lookups with MLP based classification. 
The presence of trainable embeddings help in mapping the input data from the raw input space to a latent space, which is observed to improve the classification performance.
Given our use-case, there are two main requirements we need to satisfy.
First, we need our network to have high accuracy in predicting the best runtime configuration which maximizes performance.
Second, given that the recommendation network needs to be queried at runtime, the network should be small keep the inference latency and implementation costs low.
In our use case, the recommendation inference for a given layer is run concurrent to the execution of a previous layer whenever possible. 
Lower inference latency therefore moves the recommendation step out of the critical path.
Moreover, a smaller network has fewer computation and storage requirements and hence minimizes the overheads.
Honoring these requirements, we propose a network as depicted in \autoref{fig:adaptnet-config-arch-wide}(f).
The network, called \recnet, is simple, where we lookup the embedding entries for the input features, and then use a classifier with single hidden layer with 128 nodes and softmax activation at the output.
%

\textbf{Training, Performance, and Generalizability.}
To train our recommendation network we use one Titan RTX GPU with 84 SMs. 
When training on the dataset for $2^{14}$ MAC based RSA, for 30 epochs with a mini-batch size of 32, it takes about an hour to converge. 
\autoref{fig:adaptnet-performance}(a) shows the accuracy progression as the training proceeds.
We obtain a high accuracy of 95\% of the test dataset of 200K points, which is compared against other classifiers in \autoref{fig:adaptnet-config-arch-wide}(e). 
%
We also test the robustness of our design by generating similar datasets of 2M points 
each for RSA's with varying number of MAC units (eg $2^{12}$, $2^{13}$ etc). 
The aim is to test the performance of different \recnet~ with different output configuration space.
In \autoref{fig:adaptnet-performance}(b) we plot the test accuracies obtained for each such \recnet~ trained for 30 epochs with 90:10 training-testing split. Please note that the data points in test datasets are unknown at training time. 
We observe that the networks all achieve high accuracies over 90\%. 
To distinguish the \recnet's among themselves we use the size of the configuration space as a suffix. 
For example, the design space of $2^{14}$ MAC has 858 possible configuration, therefore we call the corresponding network \recnet-858.

\subsection{Alternatives to \recnet}
Memoization, in form of caching is one alternative to \recnet{} to attain constant time configuration lookup. 
However, caching only works for a limited number of previously computed workloads. 
For any workload which does not hit in the cache, search has to be performed at runtime. 
The large configuration space of \ra{} as depicted in  \autoref{fig:adaptnet-config-arch-wide}(a) makes it a non scalable solution. 
One the other hand, \recnet{}, owing to learned parameters, can generalize configuration recommendation to any query having workload dimensions generated from the distribution of its training dataset.

\insertFigure{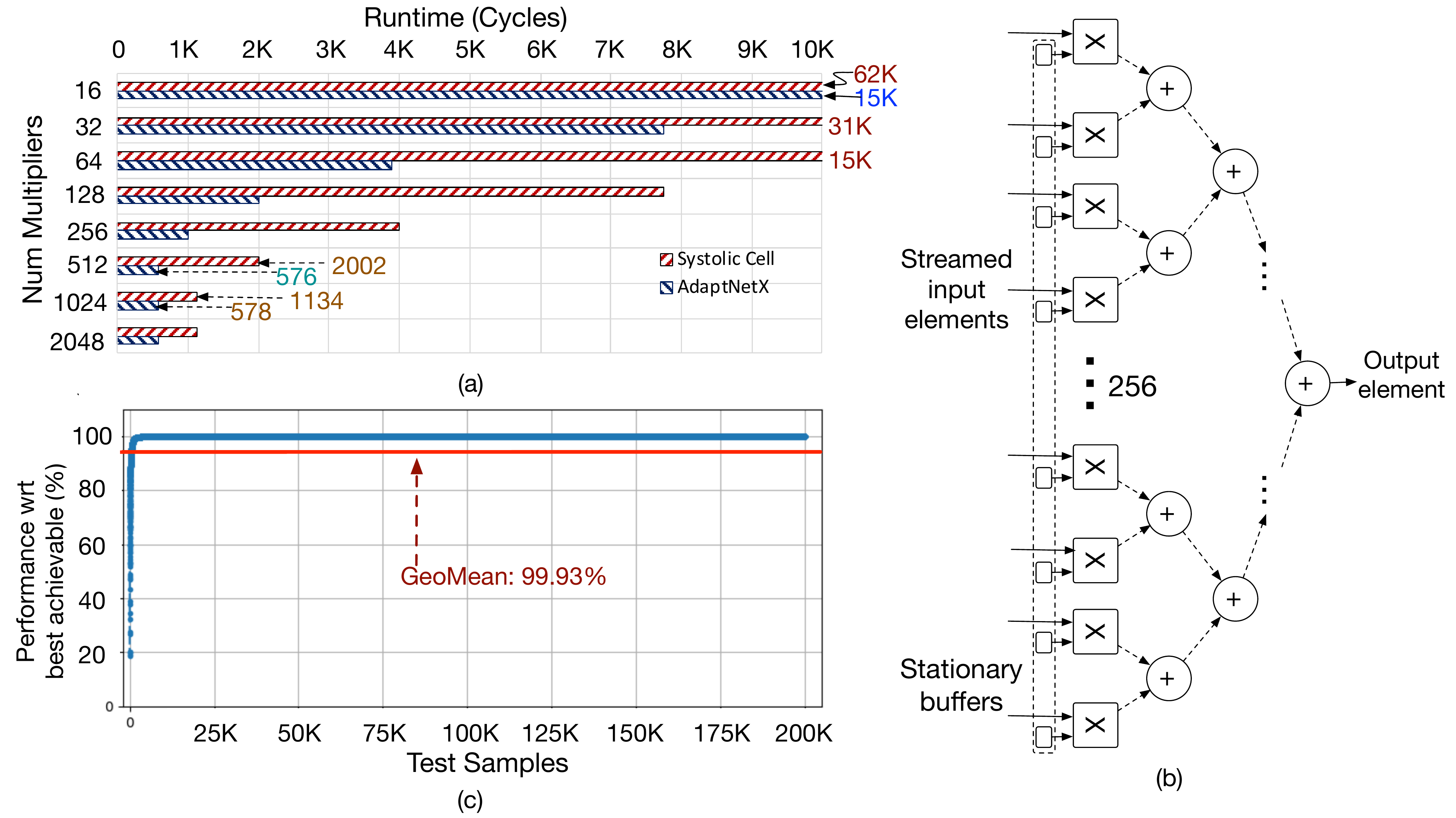}{
    (a) Cycles needed to run \recnet -858 on
     an array of \syscell s and on the custom hardware unit (\core) as a function of number of multipliers.
    (b) Architecture of the custom 1-D unit hardware for \core
    (c) Relative performance of the configurations predicted by \recnet-858 for \mys~ for $2\times10^{5}$ test samples when compared to the runtime of best possible configurations
}

\section{\textbf{\textsc{Self Adaptive Reconfigurable Arrays}}}


By coupling \recnet{} with a reconfigurable array, we can create a self adaptive system which can be conceptually viewed as a combination of two units, a Self Adaptive unit (SA), and a Reconfigurable Array (RA) unit as shown in \autoref{fig:sara-concept}.
The SA unit encompasses the software and hardware components which recommend the optimal configurations. 
The RA unit is the hardware unit capable of flexibly configuring to the recommended configurations and hence run the workloads.
It is worth pointing out that this design class is not specific to a reconfigurable core for running GEMM workloads.
Instead any Coarse Grained Reconfigurable Array (CGRA) unit, configurable at runtime, can be augmented with a suitable SA, to ensure optimal performance.
We believe this results in a new class of designs, which we name Self Adaptive Reconfigurable Array (SARA).

\vspace{-2mm}
\subsection{Hardware to run \recnet}
In the context of our use case, an intuitive option is to allocate a few \syscell s from the main array to run \recnet. 
However, this choice will lead to either fewer MAC units left for the actual workloads, or to allocate additional \syscell s for \recnet~ leading to an additional overhead. 
An alternative to adding more \syscell s will be to add a custom hardware dedicated for running \recnet.
We explore both the \syscell~and custom hardware options below for \recnet-858.


\textbf{\recnet{} Runtime on \syscell s.}
\autoref{fig:adaptnet-x-case}(a) shows the cycles required for a single inference of the \recnet ~as a function of multipliers used in $4\times4$ \syscell ~based array.
Understandably, the runtime decreases proportional to the increase in number of multipliers as we increase the number of \syscell s, achieving the best runtime of 1134 cycles when using 1024 multipliers or 64 cells. 
When both the workloads and the recommendation engine is run on a same array; for a TPU equivalent machine with $2^{14}$ MAC units, about 6.25\% of the array needs to be allocated for running the \recnet. 
Another choice could be allocating more hardware resources in terms of extra 64 \syscell s dedicated to run the recommender network. 
However, given that \recnet~ has exclusively dense layers processing the embedding lookups, a systolic execution turns out to be sub-optimal.

\textbf{\recnet{} Runtime on \core.} We found a custom design tuned for \recnet~ layer parameters to be more efficient.
For efficient execution of the dense layers, we chose a 1-D multiplier unit with a binary tree based reduction as shown in \autoref{fig:adaptnet-x-case}(b). 
We found Input stationary (IS) dataflow to be the most performant for our use case. 
In this mapping the elements of the input vector is buffered near the multipliers, while elements of the weight matrix are streamed through to generate one output element/partial sum, with a sustained throughput of 1 element per cycle.
Throughput can be further increased by adding more such 1-D units.
We name the custom core with one or more such 1-D units as \core.
In \autoref{fig:adaptnet-x-case}(a) we depict the variation of runtime of \recnet~ inference on \core~ with two 1-D units as a function of multipliers. 
We find the 512 multipliers result in best runtime of 576 cycles, when running \recnet{} for $2^{14}$ MAC unit \syscell{} design. 
We also examine the cost of misprediction of \recnet{} in \autoref{fig:adaptnet-x-case}(c), where we plot the runtime of the predicted configurations from \recnet-858 normalized to best possible runtime. 
We see that most mispredictions are benign and only a few misprediction lead to catastrophic performance losses, leading to a geometric mean of 99.93\% of the best possible performance.

\vspace{-2mm}
\subsection{\mys{} Accelerator}
\mys~ is constructed by augmenting the $2^{14}$ MAC \ra{} unit, laid out as $32\times32$ grid of \syscell s, with \core ~running \recnet-858 (see \autoref{fig:sara-system}).
We chose this configuration as it has the same compute as the TPU v2, and the $4\times4$ \syscell ~size works the best for our workloads (see \autoref{subsec:sim-eval}).
Since each row and column in this configuration has 31 bypass links and one link to MAC, each buffer is constructed as a collection of 1024 1KB banks.

\textbf{Real-time Reconfiguration.}
The \core ~uses an additional SRAM bank of 512KB to store the embedding table and the weight matrices for \recnet-858. 
Each configuration corresponds to a 3968 bit vector which sets the bypass muxes, once the layer is ready to be mapped.

\insertFigure{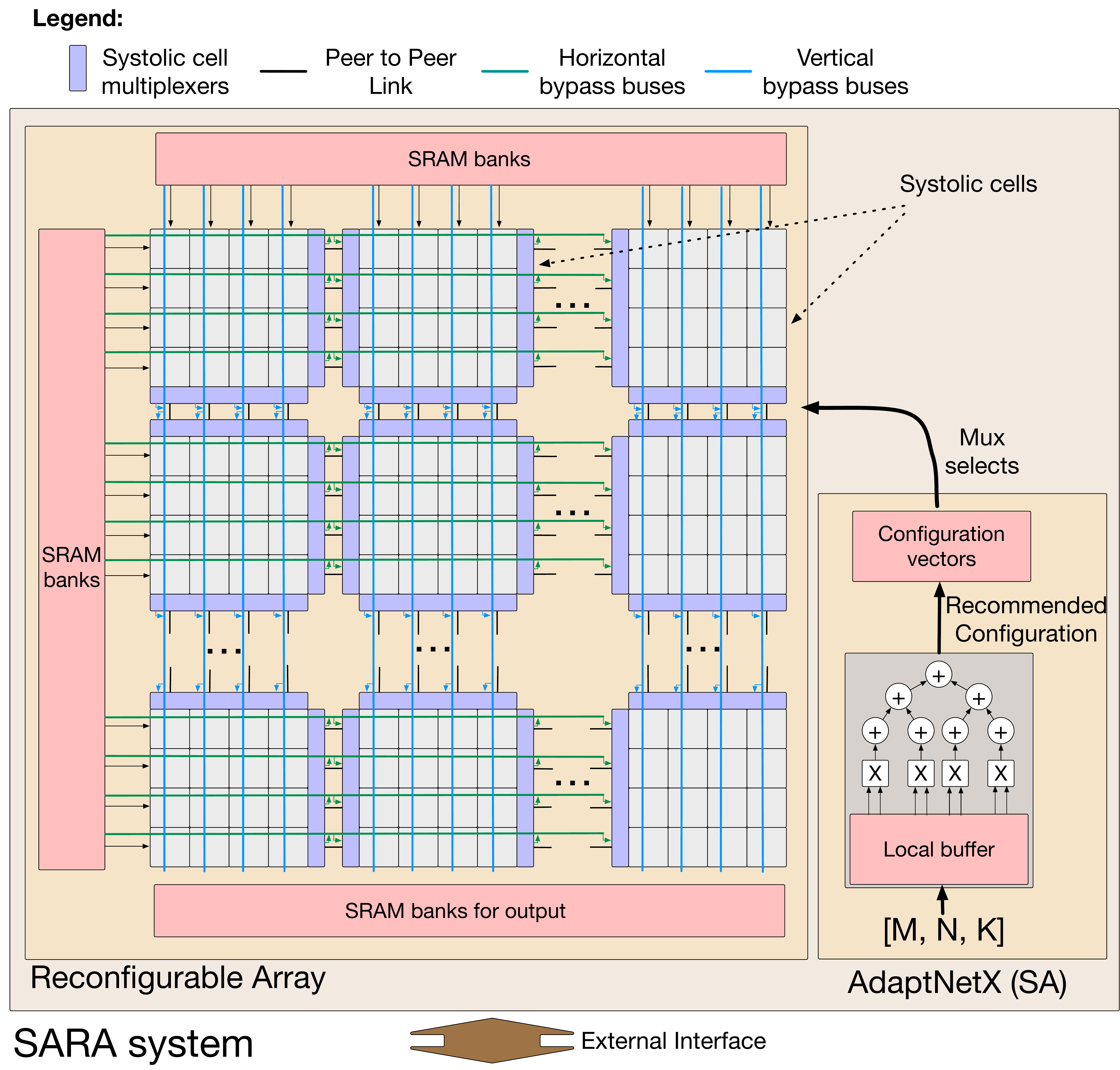}{Schematic of \mys{}, an instance of a SARA accelerator.}

\vspace{-2mm}
\section{Evaluations}
\label{sec:eval}

We evaluate \mys~ in two settings. 
To capture the merits of the architecture, we present results obtained from simulation. While the implementation aspects are captured by reporting PPA number obtained from Place-and-Route (PnR).


\subsection{Architectural evaluations}
\label{subsec:sim-eval}

\insertWideFigure{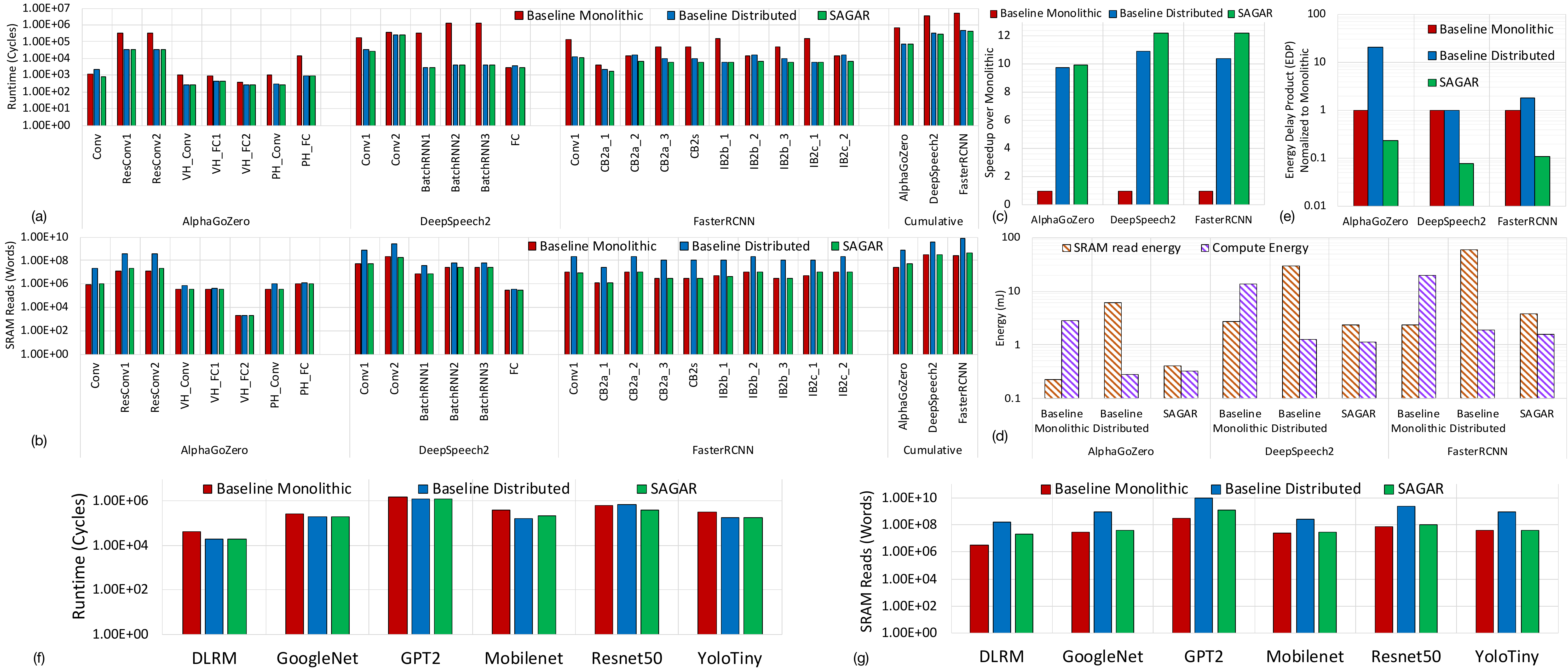}{
    (a) Simulated runtimes for monolithic $128\times128$ baseline, distributed 1024 $4\times4$ baseline, and \mys{} for layers in AlphaGoZero, DeepSpeech2, and first 10 layers of FasterRCNN 
    (b) SRAM reads for the same workloads for \mys{} and baseline configurations
    (c) Speedup of \mys{} and distributed baseline as compared to the monolithic baseline
    (d) Energy consumption breakdown for our workloads in \mys{} and baselines
    (e) Energy delay product(EDP) of \mys{} and baselines, normalized to EDP for monolithic baseline
    (f-g) Sensitivity analysis on various networks for runtime and SRAM reads
    }

\newcommand{\blD}[0]{\begin{tabular}{l} Dist. 4x4 units \\ (Baseline)  \end{tabular}}
\newcommand{\blM}[0]{\begin{tabular}{l} Mono. 128x128 \\ (Baseline)  \end{tabular}}

\begin{table}[t] 
\centering 
\setlength{\abovecaptionskip}{3pt}
\setlength{\belowcaptionskip}{0pt} 
\caption{\small Table depicting the architectural configuration of distributed systolic array based systems, monolithic systolic array baseline, and \mys 
} 
\resizebox{\linewidth}{!}{
\scriptsize \begin{tabular}{lllll}

\hline  Name &  Num Units &  MAC/unit & \begin{tabular}[c]{@{}l@{}} Banks per \\ SRAM buffer\end{tabular} & \begin{tabular}[c]{@{}l@{}} Capacity per \\ SRAM bank \end{tabular}\\

\hline
Dist. 4x4 units
& 1024 & 16 & 4 & 256 B \\
(Baseline) &&&&\\
\\

\vspace{1mm}
Dist. 8x8 units & 256	& 64 & 8 & 512 B \\

\vspace{1mm}
Dist. 16x16 units & 64 & 256 & 16 & 1 KB \\

\vspace{1mm}
Dist. 32x32 units & 16 & 1024 & 32 & 2 KB \\

\vspace{1mm}
Dist. 64x64 units & 4	& 4096 & 64	& 4 KB \\

Monolithic 128x128
& 1	& 16384	& 128 & 8 KB \\
(Baseline) &&&&\\
\\

\textbf{\mys} & \bf 1 & \bf 16384 & \bf 1024 & \bf 1 KB \\

\hline 

\end{tabular}
}
\label{table:baseline-config} 
\end{table}

\textbf{Methodology.}
For our architecture level studies we chose to use SCALE-Sim \cite{scalesim-arxiv}.
SCALE-Sim is a cycle accurate simulator for systolic array, which generates per cycle data accesses to and from various memories.
This enables us to estimate and compare performance, energy consumption, power etc. of systolic array based components to a certain degree of accuracy.
We created in-house scripts to generate SCALE-sim input files to perform the workload partitioning for the configurations recommended by \recnet-858.


\textbf{Workloads.}
For our evaluations we choose FasterRCNN\cite{fasterrcnn}, DeepSpeech2\cite{deepspeech2}, and AlphaGoZero\cite{alphagozero}, as our workloads as a representative of convolution neural networks, language modelling network, and DNNs for reinforcement learning respectively. 
\autoref{fig:sim-eval-combined}(f-g) shows our sensitivity analysis 
using a few other well known networks.

\textbf{Baselines.}
We chose a $128\times128$ monolithic systolic array and distributed array of 1024 $4\times4$ arrays as our baselines as depicted in \autoref{table:baseline-config}.
Both the arrays have same number of MAC units as TPUv2. Each array in distributed configuration resembles the tensor cores in Nvidia GPUs. 
Both that baselines have the same total SRAM memory capacity of 3MB divided into buffers for staging two operand and one output matrix.

\textbf{Performance Analysis.}
We model both of the baseline systems and \mys~in SCALE-Sim and compare the performance for our workloads. 
In \autoref{fig:sim-eval-combined}(a) we depict the cycles taken to run all the layers in AlphaGoZero, DeepSpeech2, and the first 10 layers of FasterRCNN networks.
Among the baselines, the distributed configuration mostly results in faster runtime owing to higher mapping flexibility.
However \mys, owing to reconfigurability is capable of matching the better baseline configuration.
Naturally, this flexibility leads to lower aggregated runtime for \mys ~than either of the baselines.
We see this trend generalizing in \autoref{fig:sim-eval-combined}(f) as well.

\begin{table}[t] 
\centering 
\setlength{\abovecaptionskip}{3pt}
\setlength{\belowcaptionskip}{0pt} 
\caption{Dimensions for the synthetic GEMM workloads}
\resizebox{\linewidth}{!}{
\scriptsize \begin{tabular}{c|cccccccccc}

  & G1  & G2  & G3  & G4   & G5   & G6  & G7  & G8  & G9   & G10 \\
\hline
M & 128 & 256 & 512 & 1024 & 2048 & 128 & 256 & 512 & 1024 & 2048\\ 
K & 128 & 256 & 512 & 1024 & 2048 & 64  & 64  & 64  & 64   & 64  \\ 
N & 128 & 256 & 512 & 1024 & 2048 & 64 & 64  & 64  & 64   & 64  \\
\hline
\\
  & G11 & G12 & G13 & G14  & G15  & G16 & G17 & G18 & G19  & G20 \\
\hline
M & 64  & 64  & 64  & 64   & 64   & 64  & 64  & 64  & 64   & 64 \\
K & 64  & 64  & 64  & 64   & 64   & 128 & 256 & 512 & 1024 & 2048 \\
N & 128 & 256 & 512 & 1024 & 2048 & 64  & 64  & 64  & 64   & 64 \\
\hline
\end{tabular}
}
\label{table:gemm-workloads} 
\end{table}

\insertFigure{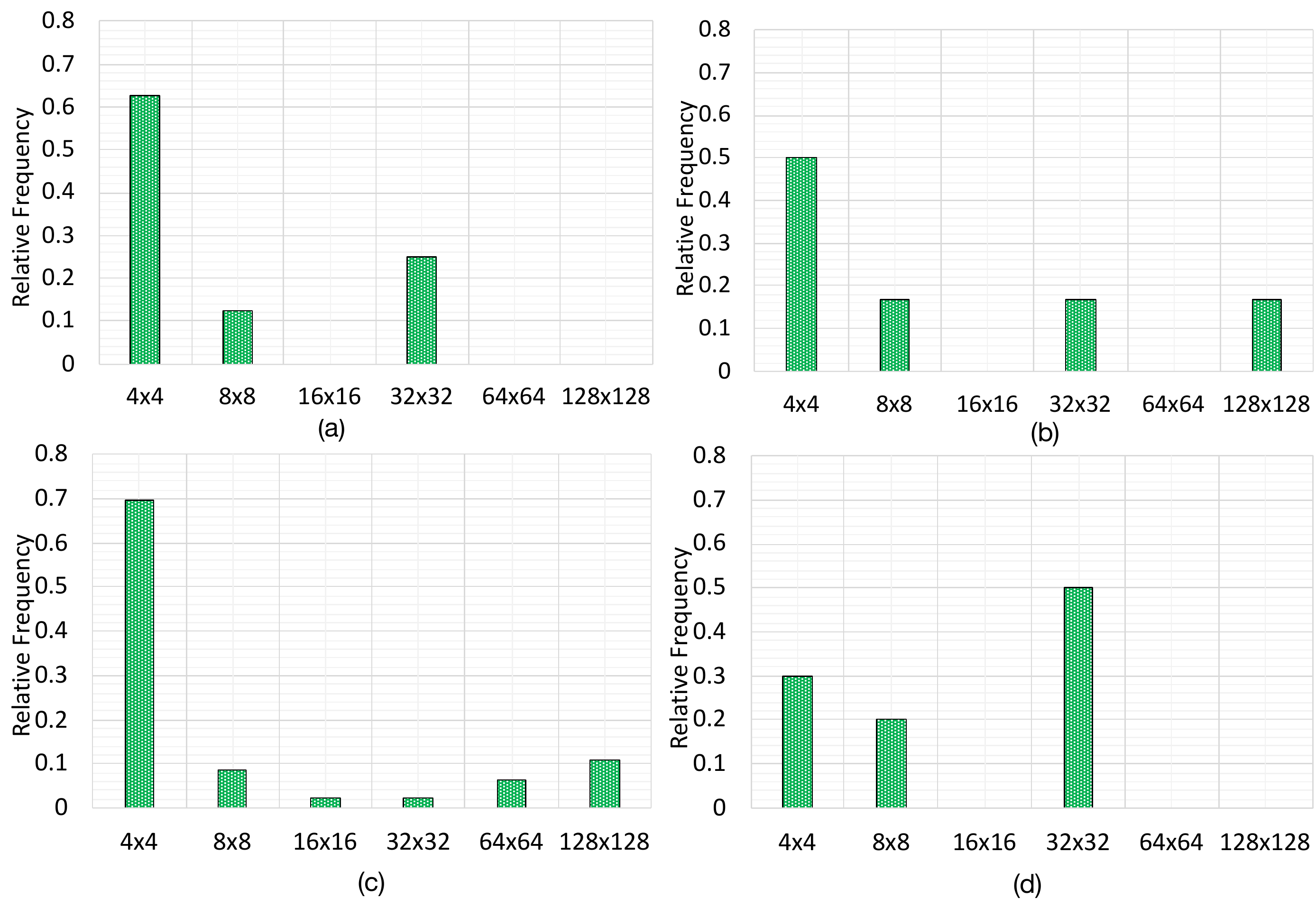}{
    Distribution of favorable array sizes for a 16384 MAC distributed system which attain the lowest runtime when run for each layer in 
    (a) synthetic GEMM workloads
    (b) AlphaGoZero, 
    (c) DeepSpeech2, and 
    (d) FasterRCNN. 
}


\textbf{\syscell{} Design Space Exploration.}
\mys ~is also capable of realizing configurations which are out of scope of either of baselines.
This allows \mys ~to achieve higher performance than both the baselines on certain layers.
For example, consider the synthetic GEMM operands depicted in \autoref{table:gemm-workloads}. 
\autoref{fig:best-config-distribution}(a) depicts the histogram of the best configuration for these layers obtained from simulation.
The layers favouring $8\times8$ or $32\times32$ configurations constitute about 40\% of the set. 
Neither of these configurations can be realized a fixed array configuration like the baselines.
In \autoref{fig:best-config-distribution}(b,c,d) we show the histogram of a similar experiment conducted on our DNN workloads.
For these specific workloads, the $4\times4$ configuration works the best for majority of the layers. 
This observation also explains our findings in \autoref{fig:sim-eval-combined}(a) on why \mys's performance is identical to the $4\times4$ baseline.
Nevertheless, for layers which favor configurations like $8\times8$, $32\times32$ etc. \mys ~will lead to lower runtime than both the baselines. 
This is depicted by \autoref{fig:sim-eval-combined}(c), where we see that \mys{} achieves about $>10\times$ speedup over monolithic when distributed configurations are preferred. While in cases where monolithic is preferred it runs faster than both the baselines.

\textbf{SRAM reads and Energy efficiency.}
%
In general, due to the loss of reuse, distributed configurations with smaller array sizes have more SRAM reads resulting in lower energy efficiency.
We observe this trend in action in \autoref{fig:sim-eval-combined}(b) where we depict the number of SRAM reads performed for layers when running our workloads on the two baselines and on \mys.
The distributed $4\times4$ system has much higher number of reads as compared to \mys ~and the monolithic baseline.
In \mys ~this efficiency loss in reuse is mitigated by using bypassing links. 
As shown in \autoref{fig:sim-eval-combined}(b), across all layers in our workloads, \mys ~incurs SRAM reads close to that in the monolithic baseline. 
In the case of DeepSpeech2, \textit{\mys}, owing to efficient mapping, incurs reads even fewer than that of the monolithic baseline.
Similar trends are also reflected in other networks as well (\autoref{fig:sim-eval-combined}(g)).
%
To further quantify the efficiency of \mys, we estimated the energy spent by the three configurations on the workloads by taking into account the cycle counts and the SRAM reads and scaling the counts by typical energy consumed per operation computed from RTL PnR flows.
For all the workloads, the wire energies calculated using 100 fJ/bit-mm at 14nm \cite{dally2020domain}, come to be about 0.1\% (maximum being 0.11\% or 0.8uJ in AlphaGoZero), which is negligible.
%
In \autoref{fig:sim-eval-combined}(d) we plot the energy consumed for the three workloads on the baselines and \mys. 
We observe that for workloads amenable to monolithic array (ie. FasterRCNN and DeepSpeech2), \mys's energy consumption is almost identical to the monolithic baseline.
The distributed baseline on the other hand consumes an order of magnitude higher energy for all the three workloads, while supporting the same mapping configurations as \mys.
The difference in energies are a direct consequence of utilization. 
Since fine grained power or clock gating is impractical, the arrays with poor utilization consume same amount of power as the arrays with better utilization. However, these arrays take longer to complete resulting in higher energy consumption.
%
For AlphaGoZero, which favours a distributed configuration, \mys ~consumes about 20\% of the energy consumed by the monolithic baseline, while almost one order of magnitude lower than that of the distributed baseline.
%
\autoref{fig:sim-eval-combined}(d) also shows that \mys's energy consumption for SRAM is close to that of consumed by the monolithic array for all the three workloads. 
The computation energy consumption in \mys ~equivalent to the better of the two baselines. 
The combined effect of improved latency and reuse is perhaps better represented by the energy-delay product (EDP) depicted by \autoref{fig:sim-eval-combined}(e).
In this figure we plot the EDP for \mys ~and the two baselines normalized to the values corresponding to the monolithic configuration. 
We observe that \mys ~results in about 92\% to 80\% less EDP compared to the monolithic baseline. 
This further demonstrates the efficiency of our proposed architecture, resulting from preserving reuse while simultaneously decreasing latency due to improved mapping.


\vspace{-3mm}
\subsection{Implementation evaluations}
\label{subsec:eval-impl}

\insertWideFigure{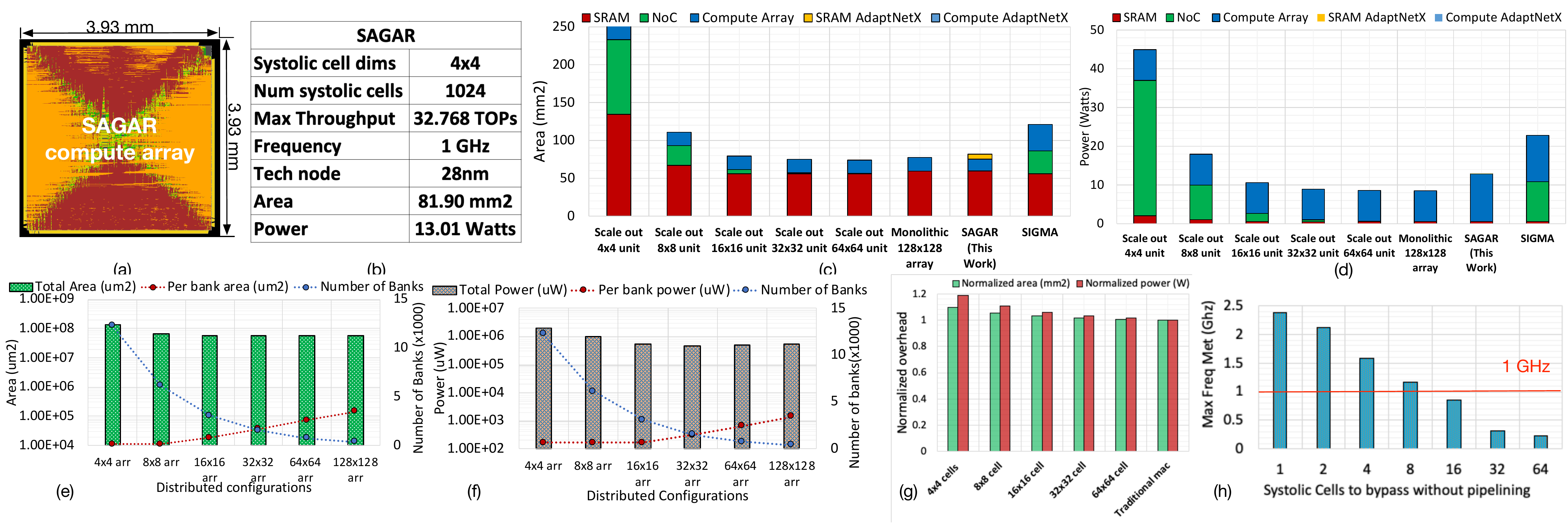}{
    Design-space exploration and final architecture of \mys.
    (a) The post PnR floor-plan diagram of \mys 's compute array, 
    (b) A table detailing architecture configuration of \mys, the implementation parameters, and post PnR area and power of \mys. 
    (c) The comparison and breakdown of post synthesis area for distributed systolic array based designs, the monolithic systolic baseline, \mys, and SIGMA 
    (d) The corresponding breakdown for power consumed by various components in distributed systolic array based designs, the monolithic systolic baseline, \mys, and SIGMA 
    (e) The variation of total area footprint of SRAM banks in various distributed systolic array and monolithic configuration juxtaposed with the variation in bank sizes and the number of banks required,
    (f) A similar variation in the power consumption by the SRAM banks in distributed systolic array and monolithic configurations, and
    (g) the the area and power of a $128\times128$ array when constructed using different sized of ``\textit{systolic-cells}'' normalized to the area and power of an array constructed with traditional MAC units.
    (h) The maximum attainable frequency vs the number of $4\times4$ \syscell s to bypass at 28nm.
}

\textbf{Methodology.}
We implemented \mys ~in RTL as a $32\times32$ array of $4\times4$ \textit{systolic-cells} and ran ASIC flow till Place-and-Route (PnR) to obtain area and power.
We used 28nm library for implementing the logic.
We also implemented the SRAM buffers as a collection of 1024 1KB cells with the SAED32 education library from Synopsis, to quantify the power and area overheads, and then scaled down to 28nm equivalent by using Dennard's scaling \cite{dennard1974design}.
\autoref{fig:rebuttal_pnr_chart}(a) depicts the post PnR floorplan of \mys's compute logic.
\autoref{fig:rebuttal_pnr_chart}(b) lists the array configuration, area, and power consumption reported after PnR by synthesizing the \ra{} and memory at a operating frequency of 1 GHz.
At 32.768 TOPs (with 1 MAC being two operation) at 1 GHz \mys ~takes 81.90 $mm^2$ of real estate while consuming 13.01 W of power. \core ~consumes 8.65\% of area and 1.36\% of power.


\textbf{Baselines.}
We implement the baseline monolithic $128\times128$ systolic array and distributed $4\times4$ array in RTL.
The distributed array is implemented using 1024 identical $4\times4$ traditional systolic arrays connected together by a mesh interconnect.
We used the OpenSMART \cite{opensmart} tool to generate and synthesize the mesh topologies for these systems.

The total memory capacity of both the monolithic and the distributed configurations are kept the same at 3MB.
As discussed in \autoref{subsec:sim-eval} the monolithic array has two input operand buffer of 1MB each and an output buffer also with the 1MB capacity.
In our implementation, we opted for one bank per row or column of the array. 
This choice ensures that each incoming link to the array will have full bandwidth from SRAM provided that bank conflicts are negligible.
Therefore each buffer in the monolithic baseline is constructed using 128, 8KB banks.
For the distributed configuration, for each $4\times4$ array we end up with 1MB for each operand buffer. 
Using the same design approach as above, we end up with each buffer being constructed using 4 banks of 256 words each.
In \autoref{table:baseline-config} we extend the same design principle for designing the memory for various other cell sizes and for \mys.
In \mys, in addition to the links going directly from the SRAM to the edge MAC units of the array, we have to consider the bypass links as well.
To get full bandwidth on these links we need to consider additional buffers. 
Extending the design described in \autoref{fig:high-bw-smart-systolic-array}, each row and column of \mys ~has 31 bypass links and one link to the first MAC unit, we need 32 banks per row/column. 
Therefore each SRAM buffer is constructed with 1024, 1KB banks.

\textbf{Area Analysis.}
In \autoref{fig:rebuttal_pnr_chart}(c) we depict the break down of area overheads for SRAM buffers, mesh NoC, and the compute array for various distributed configurations, the monolithic array, \mys ~and SIGMA~\cite{sigma}.
We observe that the monolithic configuration is the most efficient in terms of area, where it is about 5$\times$ more compact than the distributed $4\times4$ array configuration. 
The breakdown suggests that the bloating in the distributed $4\times4$ configuration is caused predominantly by the Mesh NoC (contributing to 40.5\%), followed by the SRAM buffers.
\mys ~on the other hand takes about 8\% more area than the monolithic array, while consuming about 3.2$\times$ lower area than the distributed $4\times4$ configuration.
Considering both \mys ~and the distributed configuration provides same mapping flexibility, the proposed design is strictly more efficient. 

\textbf{Power Consumption.}
In \autoref{fig:rebuttal_pnr_chart}(d) we depict the post PnR power consumption for various array configuration.
The Mesh NoC stands out as the major contributor, which naturally makes the $4\times4$ distributed configuration about 5.3$\times$ more expensive than the monolithic configuration, with the NoC contributing to about 78\% of the power.
Considering the power of the compute array alone, all the systolic-array based configurations appear to consume similar power. 
We also depict the trend in power consumed by SRAM banks across various systolic-array based configurations in 
\autoref{fig:rebuttal_pnr_chart}(f).
Similar to the trends observed in area breakdown, the counter balancing affects of increasing the bank sizes and lowering of number of banks lead to similar powers across various distributed and monolithic configurations.


\ra{} however consumes about 50\% more power than the monolithic configuration, owing to the bypass links. 
However this extra cost results in achieving the same mapping flexibility of the $4\times4$ distributed configuration, which is about 3.5$\times$ more expensive.

\textbf{Scalability Analysis.} (i) \autoref{fig:rebuttal_pnr_chart}(g) we show the overhead of using smaller \syscell{} sizes in terms of area and power normalized to monolithic configuration. 
For specific use cases with relaxed requirements for flexibility larger sized \syscell s can be used to improve the implementation costs.
(ii) \autoref{fig:rebuttal_pnr_chart}(h) we depict the max frequency that can be met as a function of number of $4\times4$ \syscell s that can be bypassed at 28 nm. 
Since we target 1GHz, we need to pipeline the bypass paths by inserting flops after 8 \syscell s as we discuss in \autoref{subsec:bypasslinks}.

\vspace{-2mm}
\subsection{Comparison with SIGMA}
\label{subsec:sigma-comp}

  
\insertFigure{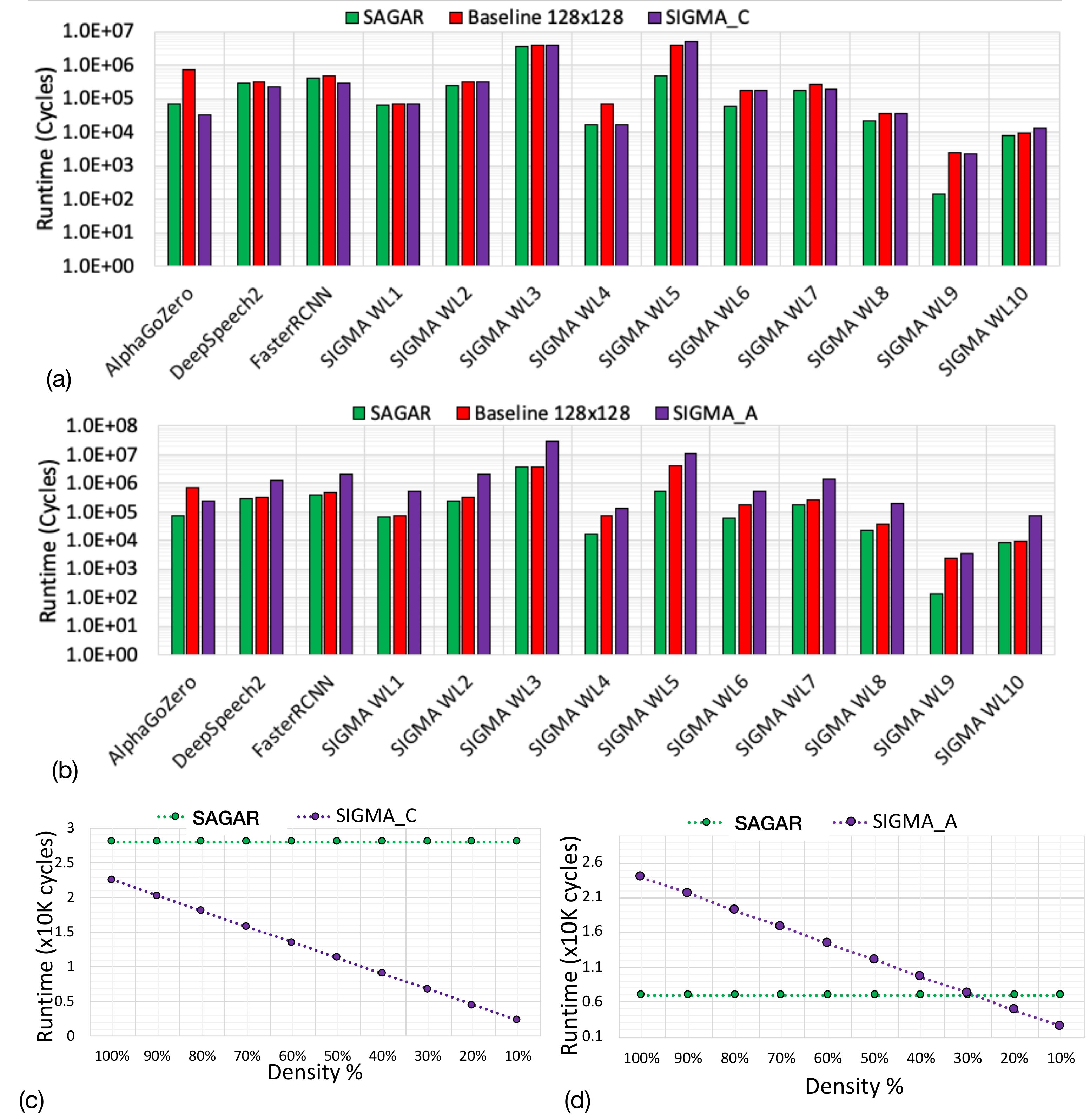}{
    Runtimes obtained for
    (a) running dense workloads for monolithic baseline, \mys, ~and \textit{compute normalized} configuration of Sigma (SIGMA\_C),
    (b) dense workloads for monolithic baseline, \mys, ~and \textit{area normalized} configuration of Sigma (SIGMA\_A); and
    (c) \mys ~and SIGMA\_C configuration by increasing levels of sparsity (decreasing density) in DeepSpeech2,
    (d) \mys ~and SIGMA\_A configuration by varying levels of sparsity in AlphaGoZero
  }

\textbf{Implementation Comparison.}
We compare the area of \mys~ with the published area and power numbers of a state-of-the-art flexible accelerator SIGMA \cite{sigma}.
SIGMA allocates a significant portion of area for NoC, which together with SRAM comprise about 80\% of the total area \autoref{fig:rebuttal_pnr_chart}(c).
In \mys, simple bypass links are used to achieve the flexibility, which saves about 30\% of the area in comparison.
From \autoref{fig:rebuttal_pnr_chart}(d), we observe that NoC is SIGMA consumes about 1.8$\times$ more power than \mys, with NoC consuming 45\% of total power.

\textbf{Performance Comparisons.}
We use the analytical model used in the original paper
\footnote{We thank the authors of SIGMA for their gracious support}
to estimate performance of SIGMA~\cite{sigma}, which accounts for the time taken to stream, compute, and add partial sums as per the functionality described in their paper.
  %
  In \autoref{fig:sigma-compare-revised}(a) we plot the simulated runtimes for  \mys, monolithic baseline, and SIGMA with equal number of MAC units (denoted as SIGMA\_C) for our representative workloads 
  and the ten layers reported in the SIGMA paper.
  %
  SIGMA\_C outperforms \mys ~in all workloads. 
  This is due to the fact that the operands are directly streamed to the multiplier over the heavy Benes network, whereas in \mys, the store-and-forward operation takes up some cycles. 
  The gap in performance further widens with the increase in sparsity as shown in \autoref{fig:sigma-compare-revised}(c).

  As SIGMA implementation takes more area than \mys, we also compare against the area normalized configuration \revised{(2734 MACs)} of SIGMA (denoted as SIGMA\_Ain \autoref{fig:sigma-compare-revised}) for fairness.
  In this case, SIGMA\_A consumes about an order of magnitude more number of cycles for each workload as compared to compute normalized configuration, therefore rendering \mys ~as the best performer (\autoref{fig:sigma-compare-revised}(b)).
  Even when considering workloads with sparse operands, SIGMA\_A is able to surpass \mys ~only at operand sparsity values above 70\% (see \autoref{fig:sigma-compare-revised}(d)).

\vspace{-2mm}
\section{Related Works}
\label{sec:related-work}

\textbf{Flexible DNN Accelerator}.
\autoref{table:related-works} depicts the standing of various such accelerators in term of native operation supported, mapping capability and flexibility.
Designs like\cite{neurocube}, \cite{flexflow}, \cite{zhang-fpga-2015}, \cite{planaria}, \cite{brainwave}, \cite{sigma}, \cite{maeri}, \cite{cascades}, \cite{zhang-fpga-2015, alwani-2016-MICRO-fused, tangram} have limited flexibility in either reconfigurability or dataflow.
The \textsc{Reconfigurable Systolic Array} enables both mapping flexibility and reconfigurability.

\textbf{Dataflow and Accelerator Design Space Search.} 
Contemporary tools \cite{scalesim-ispass}, \cite{maestro}, \cite{tetris}, \cite{timeloop}, \cite{dmazerunner} etc enable DSE by fast cost estimation or heuristics.
SARA systems like \mys{} on the other hand obtain optimized configuration at runtime in one-shot using \recnet.

\textbf{ML assisted system configuration.} 
Recent work \cite{gamma, confuciux, nautilus} show using GAs for efficient search.
RL and recommendation has been used for chip PnR \cite{mirhoseini2017device, kwon2019learning}.
AutoTVM \cite{autotvm} use ML models for cost prediction to improve compilation time.
It is worth noting that these approaches mostly enhance search for the optimal configuration, while \recnet~ replaces search.

\vspace{-2mm}
\section{Conclusions}
\label{sec:conc}

This work shows that the mapping and configuration space of reconfigurable accelerator can be learnt using ML.
We demonstrate this by developing a recommendation model called \recnet{} which learns and predicts the optimal configurations for \ra{} with high accuracy. 
\ra{} is a flexible, scalable GEMM accelerator constructed using \syscell s and pipelined bypass paths.


\bibliographystyle{IEEEtranS}
\bibliography{references}

\end{document}